\documentclass[prb,twocolumn,showpacs]{revtex4}

\usepackage{amsmath,amsfonts,amssymb,bm}
\usepackage{dcolumn}
\usepackage[final]{graphicx}
\usepackage{bm}
\usepackage{comment}

\includecomment{pdffigure}

\renewcommand{\Im}{\mathfrak{Im}}

\begin{document}

\bibliographystyle{apsrev}	

\title{Quantum corrected model for plasmonic nanoparticles:  A boundary element method implementation}

\author{Ulrich Hohenester}
\email{ulrich.hohenester@uni-graz.at}

\affiliation{Institute of Physics,
  University of Graz, Universit\"atsplatz 5, 8010 Graz, Austria}

\date{May 14, 2015}

\begin{abstract}
We present a variant of the recently developed quantum corrected model (QCM)  for plasmonic nanoparticles [Nature Commun. \textbf{3}, 825 (2012)] using non-local boundary conditions.  The QCM accounts for electron tunneling in narrow gap regions of coupled metallic nanoparticles, leading to the appearance of new charge transfer plasmons.  Our approach has the advantages that it emphasizes the non-local nature of tunneling and introduces only contact resistance, but not ohmic losses through tunneling.  Additionally, it can be implemented much easier in boundary element method (BEM) approaches.  We develop the methodology for the QCM using non-local boundary conditions, and present simulation results of our BEM implementation which are in good agreement with those of the original QCM.
\end{abstract}



\maketitle


\section{Introduction}

Plasmonics allows to manipulate light at the nanoscale and to obtain strong and very confined electromagnetic fields~\cite{maier:07,atwater:07,schuller:10,halas:10,stockman:11}.  This is achieved by binding light to coherent electron charge oscillations at metal-dielectric interfaces, so-called surface plasmons (SPs), sometimes also referred to as surface plasmon polaritons.  Recent work has addressed the question under which conditions a classical SP description in terms of a local dielectric function breaks down and quantum-mechanical corrections become mandatory.  On the one hand, at sharp edges and corners of metallic nanoparticles there is a spill-out of the electron charge distribution, due to the electron gas pressure, which leads to a nonlocal dielectric response~\cite{david:11,luo:13,mortensen:14,toscano:14} causing a blue shift of the SP resonances and a reduction of the achievable field enhancements in comparison to local descriptions~\cite{ciraci:12}.  On the other hand, for sub-nanometer gaps and sufficiently high field strengths electrons can tunnel between neighbor nanoparticles~\cite{esteban:12,david:14,esteban:15} leading to the emergence of new charge-transfer plasmons~\cite{savage:12}.  Electron transfer through larger gaps can occur in molecular tunnel junctions~\cite{tan:14}.

From the theoretical side, such quantum corrections have been modelled by introducing either modified boundary conditions or artificial materials that mimic the quantum behaviour.  In Ref.~\onlinecite{luo:13} the authors showed that a non-local dielectric response can be modelled by replacing the non-local metal with a composite material, comprising a thin dielectric layer on top of a metal with local dielectric properties.  Similarly, in the quantum-corrected model~\cite{esteban:12,esteban:15} (QCM) an artificial dielectric material is filled into the gap region, with a conductivity that reproduces the correct tunnel current between two neighbour nanoparticles.  As the tunnel current typically has an exponential dependence with respect to the gap distance~\cite{pitarke:90}, non-planar tunneling gaps must be modelled by onion-like shells of materials with different conductivities.  Different materials can be easily introduced in volume based simulation approaches, such as finite difference time domain (FDTD) simulation~\cite{yee:66,taflove:05}.

In this paper we show how to simulate tunneling effects within a boundary element method (BEM) approach~\cite{garcia:02,hohenester.cpc:12,hohenester.cpc:14b} by introducing modified non-local boundary conditions.  While the consideration of additional materials is computationally cheap in volume based simulations, it becomes computationally very demanding in BEM simulations, since usually a large number of different material layers is needed to resolve the exponential tunnel current dependence.  In contrast, the consideration of modified boundary conditions in a QCM variant has virtually no impact on the performance of BEM simulations compared to conventional ones.  We will show that both approaches, either the consideration of artificial materials or modified non-local boundary conditions, give similar results.  From a conceptual point of view, non-local boundary conditions have the advantage that they emphasize the non-local behaviour of the tunneling process and tunnel currents do not suffer from ohmic losses but are only governed by contact resistance, a finding known for a long time in the field of mesoscopic electron transport~\cite{datta:97}.


\section{Theory}

\begin{figure}
\begin{pdffigure}
\includegraphics[width=\columnwidth]{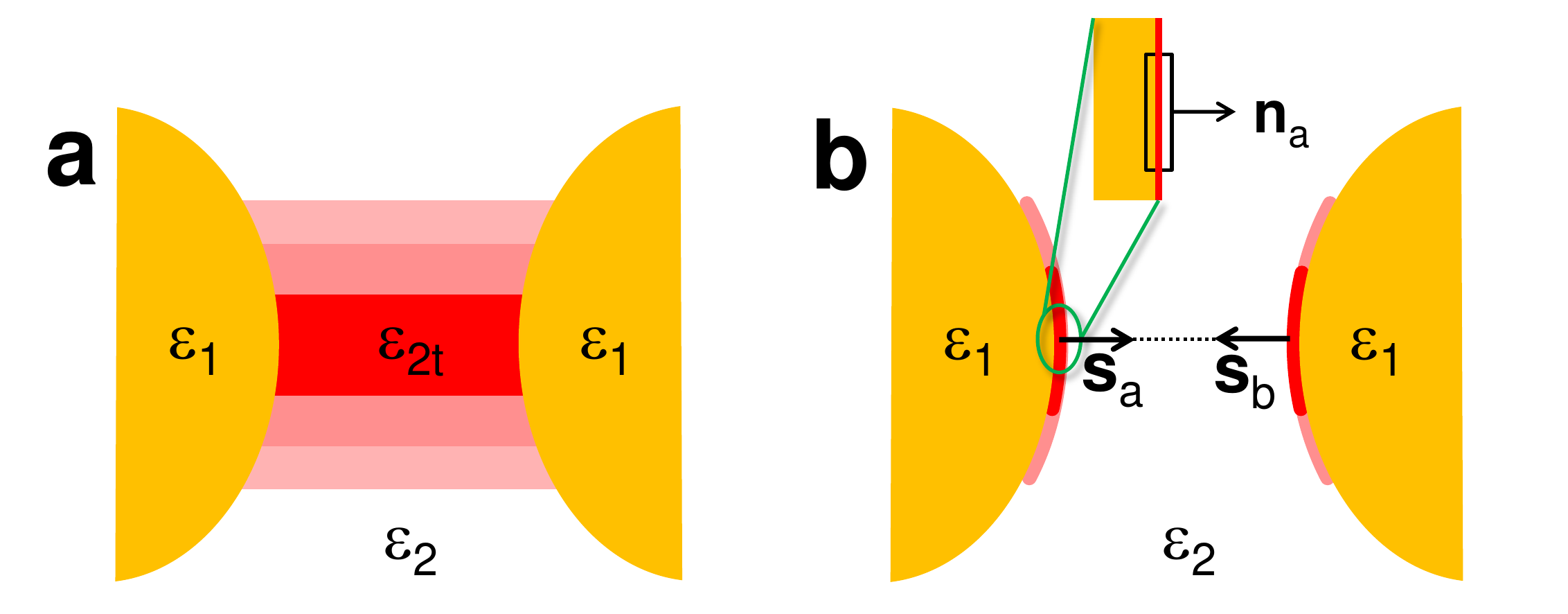}
\end{pdffigure}
\caption{(Color online) Schematics of the quantum corrected model (QCM).  (a) Volume based implementation of Esteban et al.~\cite{esteban:12,esteban:15} where artificial dielectric materials are placed inside the gap.  The conductivities of these materials are set to the gap-size dependent tunnel conductivities.  (b) Boundary element based implementation of this work, with non-local artificial boundary conditions which are chosen in order to obtain the proper tunnel current between boundary positions $\bm s_a$ and $\bm s_b$.  The inset indicates the pillbox (with outer surface normal $\hat{\bm n}_a$) over which Gauss' law is integrated to obtain the artificial boundary conditions.  For details see text.}
\end{figure}

Figure 1(a) shows the basic principle of the original QCM~\cite{esteban:12,esteban:15} (in the following denoted as volume QCM) at the example of two nanoparticles separated by a small gap of sub-nanometer size.  When an electric field $E$ is applied across the gap, a tunnel current
\begin{equation}\label{eq:current}
  J_t=\sigma_t\,E
\end{equation}
starts to flow, where $\sigma_t$ is the tunnel conductivity that can be either obtained from first principles or effective model calculations of various degrees of sophistication~\cite{esteban:12,haus:14,kaasbjerg:15,esteban:15}.  To mimic such tunnel currents, within the quantum corrected model one introduces in the gap region an effective, homogeneous medium $\varepsilon_{2t}$ with a conductivity chosen to yield the correct tunnel current (we adopt the notation of Ref.~\onlinecite{garcia:02} and denote the dielectric functions in- and outside the nanoparticle with $\varepsilon_1$ and $\varepsilon_2$, respectively).  This approach has a number of advantages:  first, it can be easily implemented in volume based simulation approaches, such as FDTD; second, the description in terms of a local current distribution guarantees that charge is conserved, i.e., the charge that leaves one nanoparticle must be transferred via the junction to the other nanoparticle.  On the other hand, the approach has a number of conceptual difficulties:  the current is subject to ohmic losses, contrary to the purely contact-like resistivity of quantum tunneling; additionally, current is not only induced by electric fields parallel the nanoparticle connection, such as one would expect for tunnel currents, but also by perpendicular fields.  In most cases of interest these are no serious shortcomings, since fields in gap regions practically always point along the nanoparticle connection, and the tunnel junction is typically so narrow that ohmic losses are of only minor importance.

We will next rephrase the QCM in terms of modified boundary conditions which are much better suited for BEM implementations.  Our starting point is Gauss' law integrated over the small pillbox indicated in Fig.~1(b),
\begin{eqnarray}\label{eq:gauss}
  \int \nabla\cdot\bm D\,d\tau&=&\oint\bm D\cdot d\bm a=4\pi\int\rho\,d\tau\nonumber\\ 
  &=&\frac{4\pi}{i\omega}
  \int\nabla\cdot\bm J_t\,d\tau=-\frac{4\pi i}\omega\oint\bm J_t\cdot d\bm a\,,\quad
\end{eqnarray}
where $d\tau$ and $d\bm a$ denote volume and surface integrations, respectively, and we have used the Fourier transformed continuity equation to relate $\rho_t$ to $\bm J_t$ (we use Gaussian units throughout).  We now make the following ad-hoc assumption for the boundary condition of the normal component of the dielectric displacement
\begin{equation}\label{eq:boundary}
  D_{2a}^\perp-D_{1a}^\perp=-\frac{4\pi i\sigma_t}\omega \,\frac{E_{2a}^\perp-E_{2b}^\perp}2\,.
\end{equation}
Here $a$ and $b$ denote the left and right nanoparticle, respectively.  The last term accounts for the charge transferred from position $\bm s_a$ to $\bm s_b$ through quantum tunneling (i.e., the loss or gain of charge in the pillbox over which Gauss' law is integrated).  Similarly to Eq.~\eqref{eq:current} we assume that the current is proportional to the tunnel conductivity $\sigma_t$ and the average of the electric field along the outer surface normal directions $\bm \hat{\bm n}_{a,b}$ [as $\hat{\bm n}_a$ and $\hat{\bm n}_b$ in the gap region are approximately antiparallel, $E_{2b}^\perp$ in Eq.~\eqref{eq:boundary} receives a negative sign].  Note that this choice is by no means unique.  We could alternatively assume $\bm J_{at}=\sigma_t(\bm E_{2a}+\bm E_{2b})/2$ or $\bm J_{at}=\sigma_t\bm E[(\bm s_a+\bm s_b)/2]$.  In all cases charge remains conserved since the current $\bm J_{at}$ leaving particle $a$ at position $\bm s_a$ is always the opposite to the current $\bm J_{bt}$ entering particle $b$, and vice versa.  However, the consideration of solely normal currents $J_{t}^\perp$ has the advantage that only the boundary condition of the dielectric displacement needs to be modified, whereas the boundary condition for the parallel magnetic field remains unaltered because of our neglect of parallel tunnel currents.

Eq.~\eqref{eq:boundary} is the central result of this work.  It replaces the consideration of artificial dielectric materials through an artificial boundary condition.  Contrary to the QCM of Esteban et al.~\cite{esteban:12,esteban:15}, our approach describes quantum tunnel as a genuine non-local process and thus does not suffer from ohmic losses in the tunnel junction.  It can be also easily extended to molecular tunnel junction by lumping all microscopic details about the microscopic tunneling process into an effective $\sigma_t$ value.  As regarding the role of normal and parallel electric fields in tunneling, both models are comparably arbitrary but could be further refined.  However, since in narrow gap regions the plasmonic nearfields preferentially point along the interparticle connection, the detailed $E^\perp$ and $\bm E^\|$ behavior of $\sigma_t$ is usually completely irrelevant.

In Appendix~\ref{sec:appendix} we show how to modify the BEM approach of Ref.~\onlinecite{garcia:02} to account for quantum tunneling, and present the working equations that can be implemented within the MNPBEM toolbox~\cite{hohenester.cpc:12,hohenester.cpc:14b}.


\section{Results}\label{sec:results}

\begin{figure}
\includegraphics[width=\columnwidth]{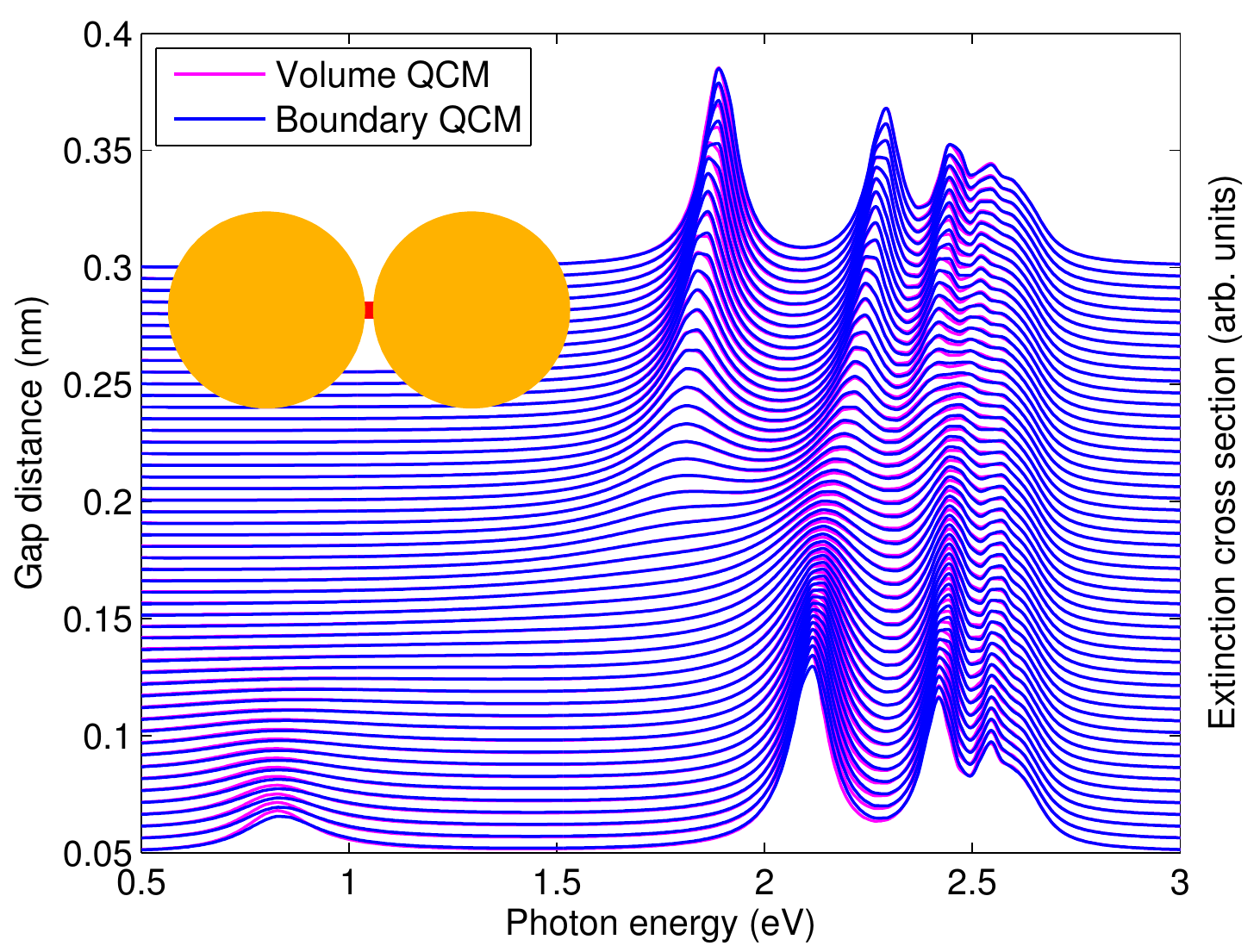}
\caption{(Color online) Comparison of volume quantum corrected model (QCM) of Esteban et al.~\cite{esteban:12,esteban:15} with the boundary QCM of this work.  We use two spheres with diameters of 50 nm and a Drude-type dielectric function representative of gold, and a \textit{single} layer of artificial tunnel material.  The light polarization is along the nanoparticle connection.  The material covers a distance range between the gap size $d_{\rm gap}$ and $d_{\rm gap}+0.2$ nm, and the artificial dielectric function is $\varepsilon_{2t}(d_{\rm gap}+0.1\,\rm nm)$.  The figure shows the gap-size dependent extinction cross section (offset for clarity, gap distance given on left axis) for the volume QCM and compares them with results of the boundary QCM.  In the latter approach, we consider quantum tunneling in the same distance window as in the volume QCM, and set the tunneling dielectric function to the same value as in the volume QCM. }
\end{figure}

We start by considering in accordance to Refs.~\onlinecite{esteban:12,esteban:15} the case of two spheres with a gap in the sub-nanometer regime.  For the dielectric function we take a Drude-type form $\varepsilon(\omega)=\varepsilon_0-\omega_p^2/(\omega^2+i\omega\gamma)$ for gold, $\varepsilon_2=1$ for the embedding medium, and 
\begin{equation}\label{eq:sigmat}
  \varepsilon_{2t}(\ell)=1+\frac{4\pi i\sigma_t(\ell)}\omega\,,\quad
  \sigma_t(\ell)=-\Im\left[\frac{\omega_p^2}{\omega^2+i\omega\gamma_p e^{\ell/\ell_c}}\right]
\end{equation}
for the tunnel material~\cite{esteban:15}.  Here $\varepsilon_0=10$, $\omega_p=9.065$ eV, $\gamma_p=0.0708$ eV, and $\ell_c=0.04$ nm, and we consider only purely imaginary conductivity corrections for the tunnel material.  These model parameters provide a good fit to experimental data \cite{johnson:72} for photon energies below 2 eV but underestimate dielectric losses above 2 eV where $d$-band scatterings set in.  Nevertheless, in this work we keep the Drude description to facilitate the comparison with Refs.~\onlinecite{esteban:12,esteban:15}.  The frequency dependence and details of $\sigma_t$ are subject of ongoing research efforts, the parametrization of Eq.~\eqref{eq:sigmat} has been motivated by static tunneling calculations including image charge effects as well as by time-dependent density functional theory calculations~\cite{esteban:15}, related work has employed theory developed for optical-assisted tunneling in the microwave domain~\cite{haus:14} or diagrammatic expansions for the ac conductance through inclusion of higher-order electron-plasmon interactions~\cite{kaasbjerg:15}.  As the primary goal of this work is the derivation and implementation of a boundary QCM  using a suitable $\sigma_t$ parametrization, we will here not further elaborate on this point.

\begin{figure}
\includegraphics[width=\columnwidth]{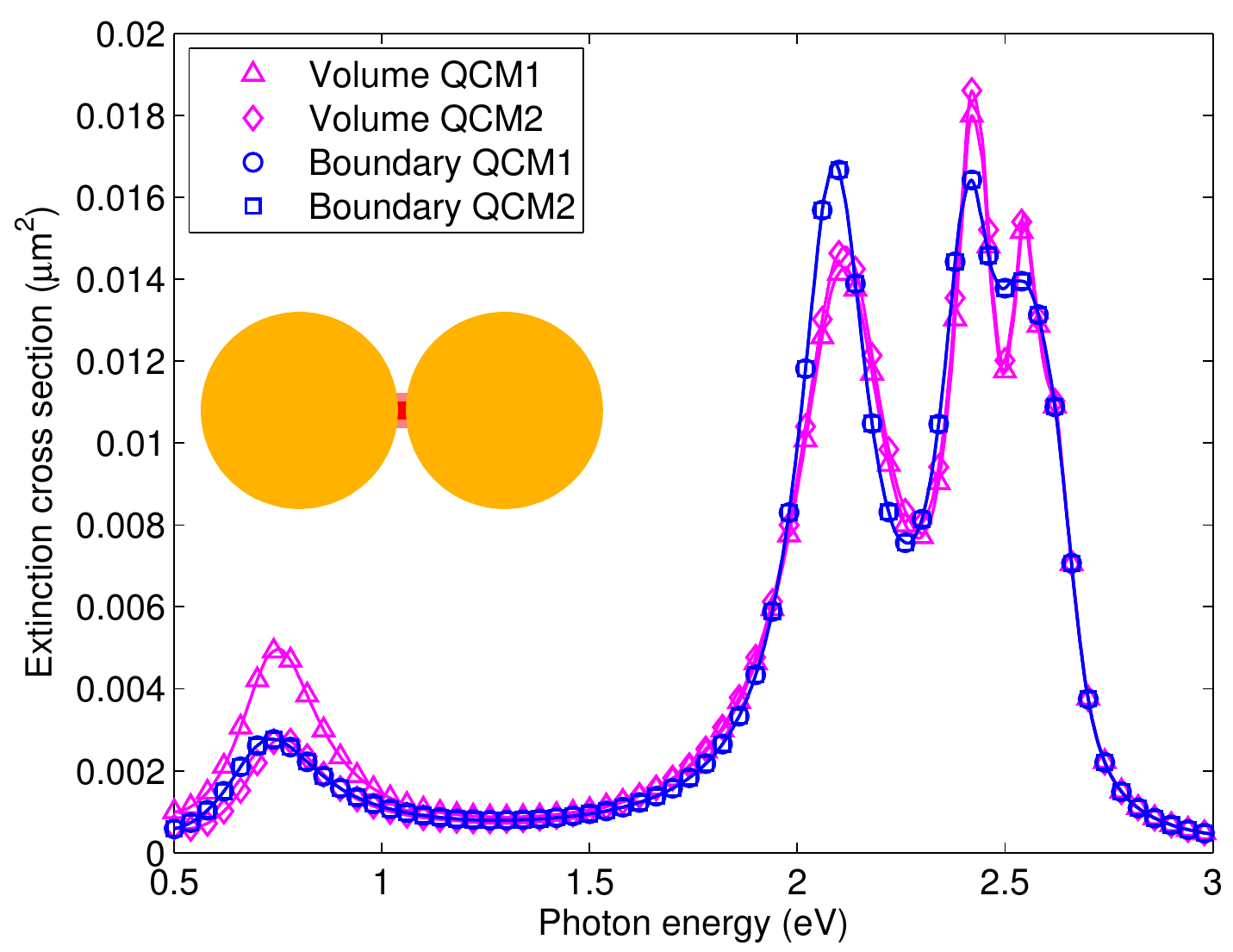}
\caption{(Color online) Volume and boundary QCM for the same spheres as in Fig.~2 and for $d_{\rm gap}=0.075$ nm.  In the volume QCM we consider an onion-like sequence of five materials $\varepsilon(\ell)$, with $\ell$ covering the region from $d_{\rm gap}$ to $d_{\rm gap}+0.4$ nm.  In the boundary QCM we use the $\varepsilon(\ell)$ values for the respective boundary element distances.  Volume QCM1 refers to the model of Ref.~\onlinecite{esteban:12} and volume QCM2 to a simulation where the light excitation and the scattered far fields are computed without the artificial materials.  Boundary QCM1 refers to simulations where opposite boundary elements of the flipped spheres are connected (with a refined mesh at the poles), and boundary QCM2 to a simulation where the respective closest boundary elements of the neighbour spheres are connected.
}
\end{figure}

Fig.~2 compares for a \textit{single} artificial tunnel material in between the two spheres (see inset) the extinction cross sections for different gap distances $d_{\rm gap}$.  The material covers the distance range from $d_{\rm gap}$ to $d_{\rm gap}+0.2$ nm and the dielectric function $\varepsilon_{2t}(d_{\rm gap}+0.1\,\text{nm})$ is evaluated at the average distance.  For the boundary QCM we use the same value for $\varepsilon_{2t}$ and connect boundary elements of the two neighbour spheres within the same distance range~\cite{comment.discretization}.  With this, we are able to compare the volume and boundary QCM directly.  As can be seen in the figure, both volume and boundary QCM give practically identical results over the entire range of gap distances where tunneling sets in.  Tunneling is evidenced by the disappearance of the lowest plasmon peak around 1.8 eV with decreasing gap distance, and the onset of the charge transfer peak around 0.8 eV.  Similarly to the extinction spectra, also the field enhancements in the gap region (not shown) computed within the volume and boundary QCM are in almost perfect agreement.  It is gratifying to see that the volume and boundary QCM models compare so well.

Next, we show in Fig.~3 results for the full QCM simulations for the same setup as in Fig.~2 and for $d_{\rm gap}=0.075$ nm.  For the volume QCM we use five layers of artificial materials, covering the distance range from $d_{\rm gap}$ to $d_{\rm gap}+0.2$ nm, and for the boundary QCM  we use for $\varepsilon_{2t}(\ell)$ the respective distances $\ell$ between opposite boundary elements.  Note that we use for both spheres the same boundary meshes with a refined discretization at one of the poles~\cite{comment.discretization}, and simply flip and displace the spheres to obtain the dimer structure shown in the inset.  Again we find good agreement between the volume and boundary QCM, although the volume QCM leads to a more pronounced exctinction peak of the charge transfer plasmon.  

We believe that this is an artefact caused by our BEM implementation of the volume QCM.  The BEM approach of Garc\'\i a de Abajo and Howie matches electromagnetic potentials at material boundaries in order to solve Maxwell's equations~\cite{garcia:02,hohenester.cpc:12}.  In this approach, an external plane wave excitation only excites materials connected with the embedding medium (in the gap region the outermost material is the last layer of artificial tunneling material) and the excitation is then passed to the inner layers through the solution of Maxwell's equations~\cite{garcia:02}.  While this causes typically no problems, it becomes computationally demanding for the inhomogeneous tunnel material which is modelled through closely spaced onion-like layers.  In our simulations we had problems to get fully converged results when increasing the number of layers, probably due to artificial reflections and transmissions of the incoming light at the layer interfaces.  When we consider the tunneling materials only in the BEM solutions and (artificially) neglect them in the light excitation (see simulation results with diamond symbols) we obtain for the charge transfer peak perfect agreement between volume and boundary QCM.  Also the (minor) differences at higher energies are probably due to implementation problems of the volume QMC within the BEM approach. 

The squares in Fig.~3 report results of a slight variant of the boundary QCM.  Here we do not connect opposite boundary elements (as one can only do for flipped nanoparticles) but connect the closest boundary elements of the two nanoparticles.  Apparently, such an approach also works for nanoparticle arrangements with a lower degree of symmetry.  As one infers from a comparison of the boundary QCM1 and QCM2 results, these two approaches are in perfect agreement.

\begin{figure}
\includegraphics[width=\columnwidth]{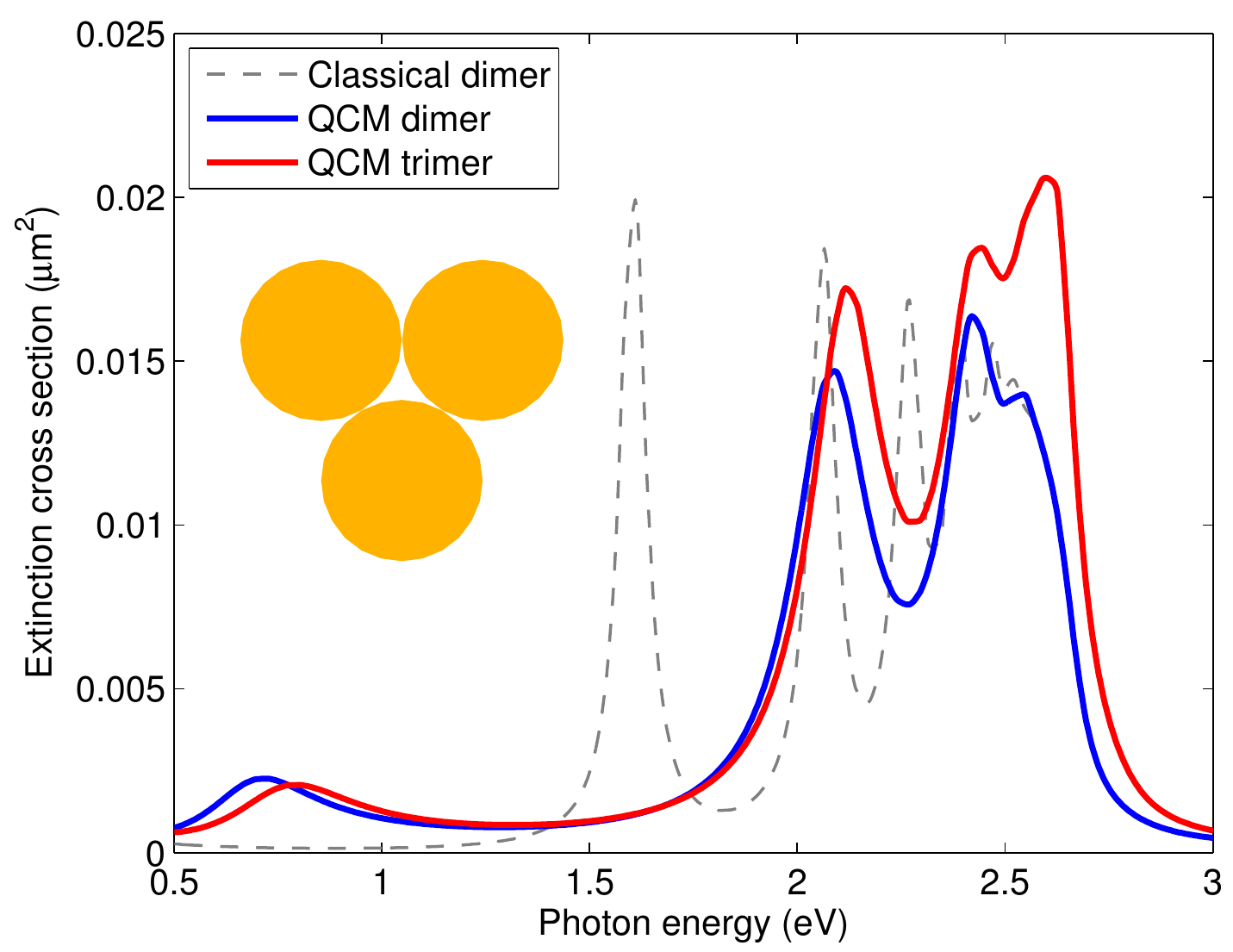}
\caption{(Color online) Extinction cross section for a dimer, using a classical electrodynamic (gray, dashed line) and a QCM simulation (blue line) with polarization along the nanoparticle connection, as well as a QCM simulation for a trimer (red line).  The sphere diameters are 50 nm and the gap distances are 0.1 nm.  For the trimer, the optical spectra do not depend on the polarization direction of the incoming light (light propagation direction perpendicular to trimer plane).}
\end{figure}

As a final example, in Fig.~4 we show results for a symmetric trimer structure consisting of three spheres, demonstrating that simulations of more complicated nanoparticles and nanoparticle arrangements can be easily performed with our BEM approach.  For the trimer structure we again observe the appearance of the charge transfer plasmon peak.  Due to the triangular symmetry, the extinction cross sections do not depend on the polarization of the incoming light (propagating perpendicularly to the trimer plane).


\section{Summary and conclusions}\label{sec:summary}

To summarize, we have presented a variant of the quantum corrected model (QCM) where tunneling is accounted for by the consideration of non-local boundary conditions.  This approach has the advantage that it emphasizes the non-local nature of tunneling and does not introduce artificial ohmic tunnel losses.  We have developed the methodology for implementing the boundary QCM within a boundary element method (BEM) approach, and have presented simulation results which have compared well with results of the original volume QCM.  Minor differences between the two approaches have been attributed to intrinsic difficulties of our BEM scheme to properly implement a volume QCM.  We believe that the volume and boundary QCM are closely related, but the availability of a different approach might be beneficial for conceptual reasons as well as for BEM implementations.

Our approach might prove particularly useful for molecular tunnel junctions with larger gap sizes.  Also supplementing the QCM through inclusion of non-local effects in the dielectric metal function, through modified boundary conditions, should be relatively straightforward.  Future work will also address the possibilities to compute the tunnel conductivities through ab-initio calculations and to submit the pertinent tunnel parameters to classical electrodynamic simulations including quantum corrections.

\section*{Acknowledgments}

This work has been supported in part by the Austrian science fund FWF under the SFB F49 NextLite and by NAWI Graz.  I am most grateful to Claudia Draxl for her hospitality during my visit at the Humboldt university of Berlin where part of this work has been performed.  Javier Aizpurua is acknowledged for helpful discussions.

\begin{appendix}

\section{}\label{sec:appendix}

Here we show how to implement the non-local quantum tunneling of Eq.~\eqref{eq:boundary} in the BEM approach of Garc\'\i a de Abajo and Howie~\cite{garcia:02} (in the following we refer to the equations of this work with a preceding G).  Importantly, we can carry over most results with the only exception of Eqs.~(G17,G18) which become modified through the nonlocal boundary condition.  

The continuity of the scalar and vector potentials $\phi$ and $\bm A$ read [Eqs.~(G10,G11)]
\begin{eqnarray*} 
  G_1\sigma_1-G_2\sigma_2 &=& \phi_2^e-\phi_1^e=\varphi\\
  G_1\bm h_1-G_2\bm h_2 &=& \bm A_2^e-\bm A_1^e=\bm a\,,
\end{eqnarray*}
where $G_1$ and $G_2$ denote the Green functions inside and outside the nanoparticle, and $\sigma$ and $\bm h$ are artificial surface and current distributions at the particle boundary which are chosen such that the boundary conditions of Maxwell's equations are fulfilled.  $\phi^e$ and $\bm A^e$ are the scalar and vector potentials of an external excitation, such as a plane wave.  For further details see Refs.~\onlinecite{garcia:02,hohenester.cpc:12}.

The continuity of the magnetic field becomes [see also Eq.~(G14)]
\begin{displaymath}
  H_1\bm h_1-H_2\bm h_2-ik\,\hat{\bm n}\left(\varepsilon_1G_1\sigma_1-\varepsilon_2 G_2\sigma_2\right)
  =\bm\alpha'\,
\end{displaymath}
with $H_{1,2}$ being the surface derivative of $G_{1,2}$ taken at the particle in- or outside, and $\bm\alpha'$ is defined through Eq.~(G15).  For the continuity of the normal dielectric displacement we get
\begin{displaymath}
  \varepsilon_1H_1\sigma_1-\varepsilon_{2t}H_2\sigma_2-ik\left(
  \varepsilon_1\hat{\bm n}\cdot G_1\bm h_1-\varepsilon_{2t}\hat{\bm n}\cdot G_2\bm h_2\right)
  ={D^e}'\,,
\end{displaymath}
with
\begin{displaymath}
  {D^e}'=\varepsilon_1\left(ik\,\hat{\bm n}\cdot\bm A_1^e-{\phi_1^e}'\right)-
         \varepsilon_{2t}\left(ik\,\hat{\bm n}\cdot\bm A_2^e-{\phi_2^e}'\right)\,.
\end{displaymath}
Here ${\phi_{1,2}^e}'$ denote the surface derivatives of the external scalar potentials, and $\varepsilon_{2t}=\varepsilon_2+(4\pi i\sigma_t/\omega)$ is a non-local dielectric function accounting for quantum tunneling, see Eq.~\eqref{eq:boundary}.  Because $\varepsilon_{2t}$ is nonlocal and connects points $\bm s_a$ and $\bm s_b$ through tunneling, it cannot be commuted with the Green functions as in the original BEM approach~\cite{garcia:02}.  Yet, the derivation of the BEM equations is not too different.

First, we use 
\begin{eqnarray*}
  G_1\sigma_1 &=& G_2\sigma_2+\varphi \\
  G_1\bm h_1 &=& G_2\bm h_2+\bm a
\end{eqnarray*}
to replace in the continuity equation (G14) of the magnetic field $\sigma_1$, $\bm h_1$ by $\sigma_2$, $\bm h_2$,
\begin{displaymath}
  \left(\Sigma_1-\Sigma_2\right)G_2\bm h_2-ik\,\hat{\bm n}\left(\varepsilon_1-\varepsilon_2\right)
  G_2\sigma_2=\bm\alpha\,,
\end{displaymath}
with $\Sigma_1=H_1G_1^{-1}$, $\Sigma_2=H_2G_2^{-1}$ and $\bm\alpha=\bm\alpha'-\Sigma_1\bm a+ik\,\hat{\bm n}\varepsilon_1\varphi$.  The continuity of the normal dielectric displacement becomes
\begin{displaymath}
  \left(\varepsilon_1\Sigma_1-\varepsilon_{2t}\Sigma_2\right)G_2\sigma_2-
  ik\left(\varepsilon_1-\varepsilon_{2t}\right)\hat{\bm n}\cdot G_2\bm h_2=D^e\,,
\end{displaymath}
with $D^e={D^e}'-\varepsilon_1\Sigma_1\varphi+ik\varepsilon_1\hat{\bm n}\cdot\bm a$.  We can use the continuity equation for the magnetic field to express the surface current $\bm h_2$ in terms of $\sigma_2$, 
\begin{equation}\label{eq:working1}
  G_2\bm h_2=\Delta^{-1}\left[ik\,\hat{\bm n}(\varepsilon_1-\varepsilon_2)G_2\sigma_2+\bm\alpha\right]\,,
\end{equation}
with $\Delta=\Sigma_1-\Sigma_2$.  Inserting this expression into the continuity equation for the normal dielectric displacement we finally obtain
\begin{eqnarray}\label{eq:working2}
  &&\Bigl[\varepsilon_1\Sigma_1-\varepsilon_{2t}\Sigma_2+
  k^2(\varepsilon_1-\varepsilon_{2t})\hat{\bm n}\cdot\Delta^{-1}\hat{\bm n}
  (\varepsilon_1-\varepsilon_2)\Bigr]G_2\sigma_2\nonumber\\
  &&\qquad=D^e+ik(\varepsilon_1-\varepsilon_{2t})\hat{\bm n}\cdot\Delta^{-1}\bm\alpha\,.
\end{eqnarray}

Equations \eqref{eq:working1} and \eqref{eq:working2} are the two working equations of our BEM approach which can be solved through matrix inversion.  Once the surface charges and currents $\sigma_2$ and $\bm h_2$ are known for a given external excitation, one can compute the electrodynamic potentials and fields everywhere else.

\end{appendix}


\begin{thebibliography}{26}
\bibitem[{com()}]{comment.discretization}
\bibinfo{note}{{ For the sphere discretization we typically use grid sizes with
  20 azimuthal angles, 5--10 polar angles for each layer of the tunneling
  material, and 20 polar angles for the remaining sphere. The ribbons of the
  onion-like tunnel materials have about 10 discretization points along the
  nanoparticle connection. We perform refined boundary element integrations
  using the MNPBEM toolbox~\cite{hohenester.cpc:12}, and checked the
  convergence of our simulations by systematically increasing the number of
  discretization points. }}.
\end{thebibliography}

\end{document}